\documentstyle[prl,twocolumn,aps,epsfig]{revtex} 

 \draft

\begin{document}

\title{Dynamic structure factor of a Bose Einstein condensate
in a 1D  optical lattice}
\author{$^{1,2}$C. Menotti, $^1$M. Kr\"amer, $^{1,3}$L. Pitaevskii 
and $^1$S. Stringari}
\address{$^1$ Dipartimento di Fisica, Universit\`{a} di Trento and BEC-INFM,
I-38050 Povo, Italy}
\address{$^2$ Dipartimento di Matematica e Fisica,
Universit\`{a} Cattolica del Sacro Cuore,
I-25121 Brescia, Italy}
\address{$^3$ Kapitza Institute for Physical Problems, ul. Kosygina 2, 
117334 Moscow, Russia}

\date{\today}

\maketitle

\begin{abstract}
\noindent
We study the effect of a one dimensional periodic potential on the 
dynamic structure factor of an interacting Bose Einstein condensate 
at zero temperature. 
We show that, due to phononic correlations, the excitation strength 
towards the first band develops a typical oscillating behaviour as a 
function of the momentum transfer, and vanishes at even multiples of 
the Bragg momentum. 
The effects of interactions on the static structure factor 
are found to be significantly amplified by the presence of the optical 
potential.
Our predictions can be tested in stimulated photon scattering 
experiments.
\\
\end{abstract}

When a Bose-Einstein condensate is loaded into an optical lattice,
its properties change in a very strong way \cite{denschlag}. 
For deep potential wells, it  even happens that the 
coherence of the sample is lost and one observes the transition
to the Mott-insulator phase \cite{jaksch,greiner}. 
Interesting phenomena occur also for low optical potential 
depth, for instance Bloch oscillations \cite{morsch} and tunneling 
effects \cite{kasevich,smerzi1,cataliotti} can be investigated in 
this regime. 
In 1D optical lattices the transition to the insulator phase is
expected to take place for very large intensities of the optical
lattice, so that there is a very extended range of parameters where 
the gas can be described as a fully coherent system.

In this Letter, we study the elementary excitations of an interacting 
Bose gas in the presence of a periodic potential and discuss how 
these states can be excited via inelastic processes using, for example, 
Bragg spectroscopy \cite{ketterle,davidson}.
To this purpose we develop the formalism of the dynamic structure
factor, a quantity directly related to the linear response of 
the system.

We will restrict ourselves to the case of a system in the presence 
of a 1-dimensional optical potential 

\begin{eqnarray}
V(z)= s\; E_R   \;{\rm sin}^2\left( {\pi z \over d} \right)
\label{Vext}
\end{eqnarray}
created by two counterpropagating laser beams. In Eq.(\ref{Vext}) 
$d$ is the lattice spacing and $s$ is a dimensionless parameter 
which denotes the intensity of the laser in units of the recoil energy 
$E_R=q_B^2/2m$. Here $q_B=\hbar \pi/d$ is the Bragg momentum 
denoting the boundary of the first Brillouin zone and $m$ is the 
atomic mass.
The inclusion of an additional harmonic potential produced, for
example, by magnetic trapping does not modify the excitation spectrum 
in a profound way, unless the wave length of the excitation is 
comparable with the size of the sample.
Along the tranverse directions we assume uniform confinement,
so that the 3D Gross-Pitaevkii (GP) equation for the ground state 
order parameter $\varphi(z)$ takes the 1D form

 \begin{eqnarray}
\left[-{\hbar^2 \over 2m} {\partial^2\over \partial z^2}
+s\;E_R \; {\rm sin}^2\left({ \pi z \over d  }\right)
+{g n d  }|\varphi(z)|^2
\right]\varphi(z)=  \nonumber     \\
={\mu }\varphi(z)\,,
\label{gpe}
\end{eqnarray}
where $n$ is the {\it average} 3D density and the order parameter 
$\varphi$ is normalized according to
$ \int_{-d/2}^{d/2}|\varphi(z)|^2 dz =1$. 
As usual  $g=4 \pi \hbar^2  a /m$ is the interaction coupling
constant fixed by the scattering length $a$.

\begin{center}
\begin{figure}
\includegraphics[width=0.95\linewidth]{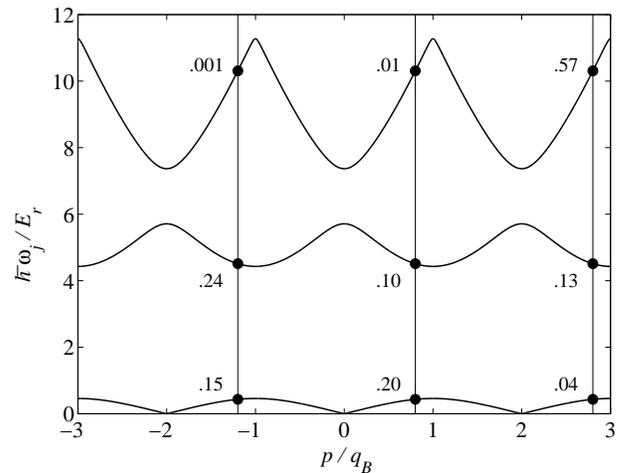}
\caption{Bogoliubov bands for $s=10$ and $gn=0.5 E_R$; 
excitation strengths $Z_j$ towards the states in the first three bands for
$p=-1.2 q_B$, $p=0.8 q_B$ and $p=2.8 q_B$.}
\label{band_10}
\end{figure}
\end{center}

The elementary excitations correspond to the solutions of the 
linearized time-dependent GP-equation and are described by the 
Bogoliubov equations

\begin{eqnarray}
\left[  
-{\hbar^2 \over 2m} {\partial^2 \over \partial z^2} 
+ s \; E_R \; {\rm sin}^2\left({\pi z \over d }\right)
- \mu + 2 gn d |\varphi|^2 \right] u_{jq} + \nonumber \\
+ gnd  \varphi^2  v_{jq} = \hbar \omega_{j}(q) u_{jq}, \label{bog_u}
 \\
\left[
-{\hbar^2 \over 2m} {\partial^2 \over \partial z^2} 
+ s \;E_R \;  {\rm sin}^2\left({\pi z \over d }\right)
- \mu + 2 gnd |\varphi|^2 \right] v_{jq} + \nonumber \\
+ gn d \varphi^{*2}  u_{jq} = - \hbar \omega_{j}(q) v_{jq},
\label{bog_v}
\end{eqnarray}
where $\varphi$ is the ground state solution of Eq.(\ref{gpe})
and the amplitudes $u_{jq}$ and $v_{jq}$ satisfy the normalization
condition 
$\int_{-d/2}^{d/2} \left[ |u_{jq}(z)|^2-|v_{jq}(z)|^2 \right]dz =1$.
The solutions $u_{jq}(z)$ and $v_{jq}(z)$ are Bloch waves: 
$u_{jq}(z)=\exp(iqz/\hbar) \tilde{u}_{jq}(z)$ where ${\tilde u}_{jq}$ is 
periodic in space with period $d$ and analogously for $v_{jq}$.
For each value of the quasi-momentum $q$,  Eqs.(\ref{bog_u},\ref{bog_v}) 
provide an infinite set of solutions $\omega_{j}(q)$, forming a band 
structure labelled with $j$ (``Bogoliubov bands'').
Due to the periodicity of the problem, the solutions of 
Eqs.(\ref{bog_u},\ref{bog_v}) with $q$ restricted to the first Brillouin 
zone and $j$ varying over all the bands, exhaust the elementary 
excitations of the system. 
Still it is often convenient to consider values of $q$ outside the 
first Brillouin zone and to treat the energy spectrum and the functions  
$u_{jq}$ and $v_{jq}$  as periodic in quasi-momentum space with 
period $2 q_B$ (see Fig.\ref{band_10}).
The intensity $s$ of the optical potential and ratio $gn/E_R$ between 
the interaction and the recoil energy are the relevant dimensionless 
parameters in terms of which we will discuss the physical behaviour 
of the system.
First numerical solutions of Eqs.(\ref{bog_u},\ref{bog_v}) in the
presence of a periodic potential were obtained in \cite{moelmer-chiofalo}.

The capability of the system to respond to an excitation probe
transferring  momentum $p$ and energy $\hbar \omega$ is 
described by the dynamic structure factor. In the presence 
of a periodic potential the dynamic structure factor takes the form

\begin{eqnarray}
S(p,\omega) = \sum_j Z_j(p) \delta(\omega - \omega_j(p)),
\label{struc-fac}
\end{eqnarray} 
where $Z_j(p)$ are the excitation strengths relative to the $j^{th}$ band
(see Eq.(\ref{strength}) below) and $\hbar \omega_j(p)$ are the 
corresponding excitation energies,  defined by the solutions of 
Eqs.(\ref{bog_u},\ref{bog_v}).
Note that $p$, here assumed to be along the optical lattice 
($z$ axis), is not restricted to the first Brillouin zone, being the 
momentum transferred by the external probe. In this respect, 
it is important to point out that, while the excitation energies
$\hbar \omega_j(p)$ are  periodic as a function of $p$, 
this is not true for the excitation strengths $Z_j$.

The dynamic structure factor satisfies important sum-rules.
The integral of the dynamic structure factor provides the static 
structure factor (non energy weighted sum-rule) 

\begin{eqnarray}
S(p)=\int S(p,\omega) d \omega.
\label{static}
\end{eqnarray}
%
As we will see later $S(p)$ is strongly affected by the 
combined presence of two-body interactions and of the optical lattice.

A second important sum-rule obeyed by the dynamic structure factor is
the model independent $f$-sum rule

\begin{eqnarray}
\int \hbar \omega S(p,\omega) d\omega = {p^2 \over 2m }.
\label{f-sum-rule}
\end{eqnarray}

Another important sum-rule is the compressibility sum-rule
(inverse energy weighted sum-rule) 

\begin{eqnarray}
\left. \int {S(p,\omega) \over \hbar \omega} \; d\omega \; \right|_{p \to 0}
= \frac{\kappa}{2},
\label{inv-sum-rule}
\end{eqnarray}
where $\kappa= [n (\partial \mu  / \partial n)  ]^{-1}$ is the
thermodynamic compressibility. 
The density dependence of the chemical potential, and hence $\kappa$,
can be obtained by solving the GP equation (\ref{gpe})  \cite{meret2}. 
The compressibility $\kappa$ is naturally expressed in terms of 
the sound velocity $c$, characterizing the low $q$ phononic behaviour 
of the dispersion law ($\hbar \omega = c q$), through the relation

\begin{eqnarray}
\kappa  = \frac{1}{m^* c^2},
\label{inv-sum-rule2}
\end{eqnarray}
where the effective mass $m^*$ differs from the bare mass because 
the Hamiltonian is not translationally invariant.
The effective mass in the presence of the external potential 
(\ref{Vext}) has been recently calculated in \cite{meret1}.

\begin{center}
\begin{figure}
\includegraphics[width=0.95\linewidth]{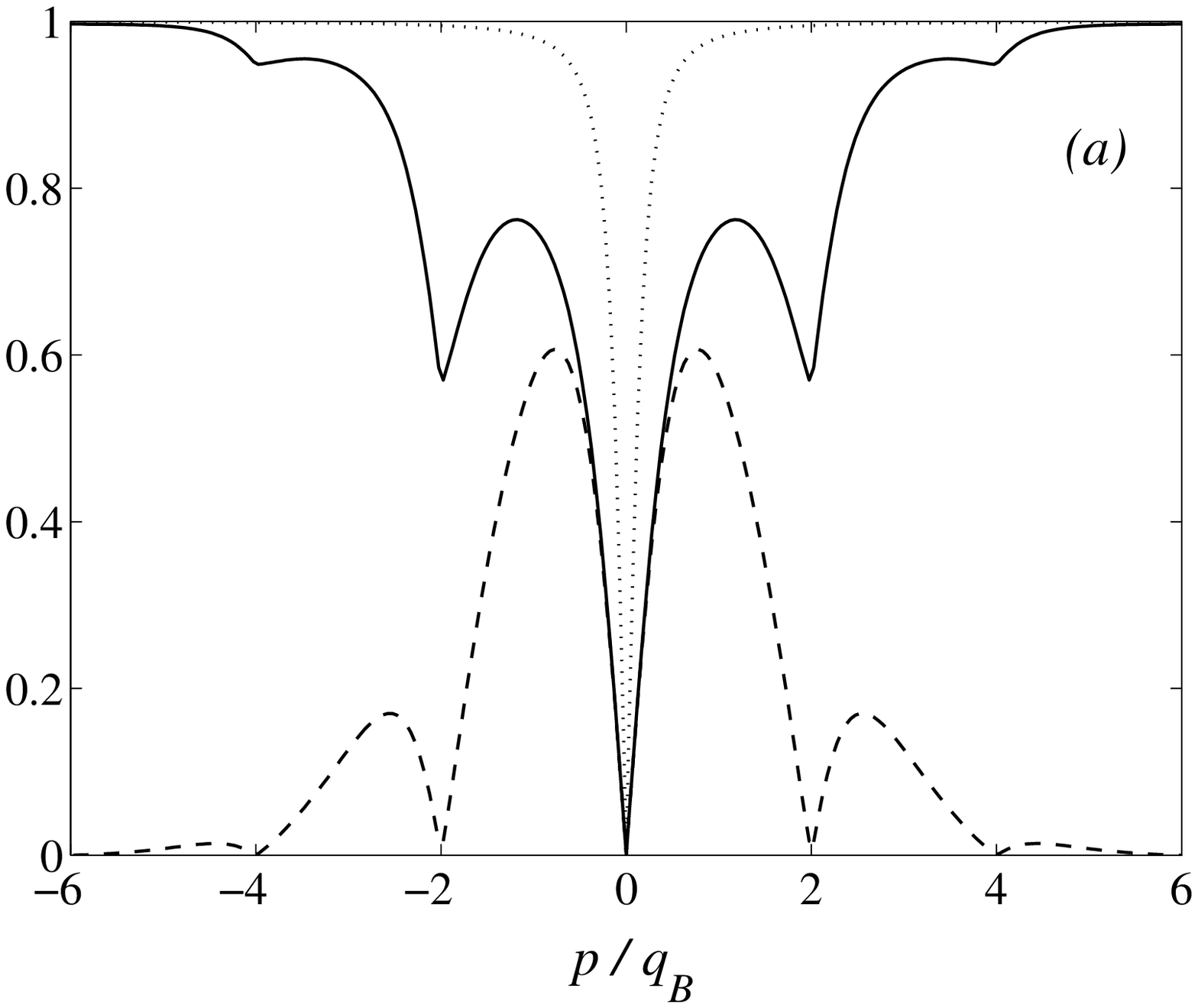}
\includegraphics[width=0.95\linewidth]{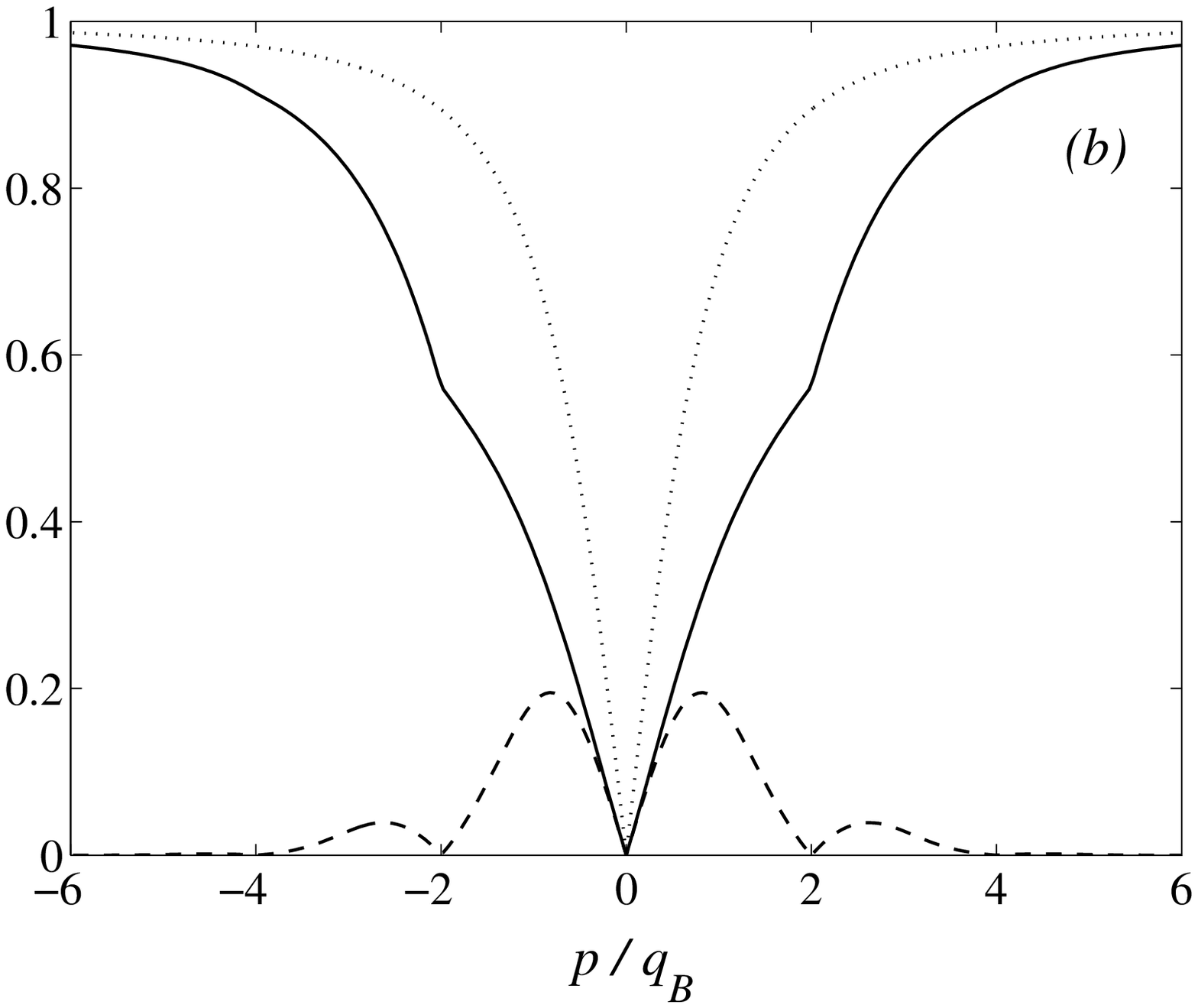}
\caption{Static structure factor (full line), $Z_1(p)$ (dashed line)
for $s=10$ and static structure factor in the uniform gas 
($s=0$, dotted line); for $gn=0.02 E_R$ (a) and $gn=0.5 E_R$ (b).}
\label{fig_Sp_10}
\end{figure}
\end{center}

In a uniform Bose gas, the sum (\ref{struc-fac}) is exhausted by 
a single mode with energy $\hbar \omega_B(p) = 
\sqrt{  { p^2 /2m}  \left({p^2 /2m}  + 2 gn \right)}$.
In this case the static structure factor obeys the Feynman relation

\begin{eqnarray}
S_B(p)=
{p^2  \over 2m \hbar \omega_B(p)},
\label{feynman}
\end{eqnarray}
where we have used the $f$-sum rule (\ref{f-sum-rule}).
For $p \to 0$ the static structure factor (\ref{feynman}) behaves like  
$|p|/2mc_B$, while the compressibility sum-rule (\ref{inv-sum-rule})
becomes $1/2m c_B^2$, where $c_B=\sqrt{gn/m}$ is the Bogoliubov 
sound velocity. 
The suppression of $S_B(p)$ at small momenta is a direct consequence of 
the phononic correlations. 
For large momenta, instead, the static structure factor (\ref{feynman}) 
approaches unity (see dotted  lines in Figs.\ref{fig_Sp_10}).
Notice that in the absence of two-body interactions, $S(p)=1$
for any value of $p$.

In the presence of the optical lattice the behaviour of the dynamic
structure factor changes in a drastic way. In particular, for a given 
value of  momentum transfer $p$, it is possible to excite several
states, corresponding to different bands (see Eq.(\ref{struc-fac}) and
Fig.\ref{band_10}). 
An important consequence is that on one hand it is possible to excite 
high energy states with small values of $p$, and on the other 
hand one can excite low energy states, belonging to the lowest band,
also with  high momenta $p$ outside the first Brillouin zone.
This behaviour introduces new possibilities in Bragg spectroscopy
experiments.

In general, the dynamic structure factor has to be calculated numerically. 
Starting from the solution of Eqs.(\ref{bog_u},\ref{bog_v}),
the excitation strengths $Z_j$ can be evaluated using the standard 
prescription (see for example \cite{griffin})

\begin{eqnarray}
Z_j(p) = \left| \int_{-d/2}^{d/2} 
\left[ u^*_{jq}(z) + v^*_{jq}(z) \right] e^{ipz/\hbar} \varphi(z) 
dz \right|^2 ,
\label{strength}
\end{eqnarray}
where $q$ belongs to the first Brillouin zone and is fixed by the relation 
$q=p+2 \ell q_B$ with $\ell$ integer.
This equation shows that, by solving Eqs. (\ref{bog_u},\ref{bog_v})
within the first Brillouin zone, one can calculate the strength
$Z_j(p)$ also for values of $p$ outside the first Brillouin zone.
In Figs.\ref{fig_Sp_10}(a,b), we show our results for two different 
choices of the interaction at $s=10$.

Let us first discuss the dynamic structure factor in the low energy 
region and in particular the behaviour of the contribution $Z_1(p)$ 
arising from the first band (dashed line).
We find that  $Z_1(p)$ exhibits characteristic oscillations
whose amplitude is suppressed at large $p$.
The zeros of $Z_1(p)$ at $p=2 \ell q_B$ ($\ell$ integer) 
reflect directly the phonon behaviour of the excitation spectrum 
which vanishes at the same values (see Fig.\ref{band_10}).

The behaviour of $Z_1(p)$ can be studied analytically in the large
$s$ limit where the tight binding approximation applies.
In this limit one can approximate the solutions of Eqs. 
(\ref{bog_u},\ref{bog_v}) in the lowest band by

\begin{eqnarray}
u_q(z)= U_q \sum_k e^{i q kd / \hbar} f(z-kd)
\end{eqnarray}
and analogously for $v_q(z)$, where $f(z)$ is a function localized near
the bottom of the optical potential $V$ at $z=0$, and $k$ labels the 
potential wells. Within this approximation the function $f$ also 
characterizes the ground state order parameter which reads 
$\varphi(z) = \sum_k f(z-kd)$.

In the tight binding approximation the dispersion law of the lowest
band takes the Bogoliubov-like form \cite{javanainen}

\begin{eqnarray}
\label{w1}
\hbar \omega(p) &=&
\sqrt{ \varepsilon(p) 
\left[ \varepsilon(p)  + 2 \kappa^{-1}\right]},
\label{t-b}
\end{eqnarray}
where

\begin{eqnarray}
\label{e1}
\varepsilon(p) &=&
2 \; \delta \; {\rm sin}^2  \left({p d \over 2 \hbar}\right).
\end{eqnarray}
In the above equations $\delta$ is the tunneling rate of particles
between two consecutive wells. The tunneling rate is related to the 
effective mass entering the compressibility sum-rule
(\ref{inv-sum-rule},\ref{inv-sum-rule2}) by 
$\delta=2 m E_R/ \pi^2 m^* $
and decreases by increasing the laser intensity $s$. 
The parameter $\kappa$ is the compressibility of the gas as emerges 
from the low momentum behaviour of the dispersion law (\ref{t-b}): 
$\hbar \omega = \sqrt{\kappa^{-1} / m^*} p$.
In the tight binding limit, one finds
$\kappa^{-1} = gnd \int_{-d/2}^{d/2} f^4(z) dz$ \cite{note}.

By approximating the function $f(z)$ with the gaussian 
$f(z)= \exp[-z^2/2 \sigma^2 ] / (\pi^{1/4}\sqrt{\sigma })$,
one finds, after some straightforward algebra, the result

\begin{eqnarray}
Z_1(p) = { \varepsilon(p) \over  \hbar \omega(p)} 
\exp\left(-{ \pi^2 \sigma^2 p^2 \over 2 d^2  q_B^2}\right)
\label{gen-feynman}
\end{eqnarray}
for the strength relative to the first band, where the width $\sigma$ 
can be calculated numerically by minimization of the ground state
energy and behaves like 
$\sigma \sim s^{-1/4} d/ \pi$ for $s \gg 1$.
Equation (\ref{gen-feynman}) reproduces with good accuracy
the numerical results obtained by solving the Bogoliubov
equations for relatively large values of $s$. It accounts for both the 
suppression of the strength at large $p$ through the gaussian term,
and the oscillating behaviour through the Feynman-like
term $\varepsilon(p) /  \hbar \omega(p)$.
The strength $Z_1$ has a maximum close to the edge of the first Brillouin
zone, where it takes approximately the value
$Z_1(q_B) \approx \sqrt {\kappa \delta / ( \kappa \delta  + 1 )}$.
This simple expression shows that  $Z_1$ is quenched both by 
increasing interactions ($\kappa \to 0$) and by increasing the optical 
potential ($\delta \to 0$).
In the cases of Figs.\ref{band_10}(a,b) one has $\kappa \delta= 0.95$
and $0.056$ respectively.
In the noninteracting case ($\kappa^{-1}=0$) one has 
$\varepsilon(p) = \hbar \omega(p)$ and the strength (\ref{gen-feynman}) 
reduces to $Z_1(p) = \exp\left(- \pi^2 \sigma^2 p^2 / 2 d^2 q_B^2 \right)$.
The comparison then clearly shows that the oscillating behaviour
of $Z_1(p)$ as well as its quenching at large $s$ are a direct
consequence of two-body interactions.
On the other hand, two-body interactions scarcely affect the strengths
towards the higher bands, provided $gn \ll \sqrt{s} E_R$.

The quantity $Z_1(p)$ could be measured in Bragg spectroscopy experiments
by tuning the momentum and the energy transferred by the scattering 
photon to the values of $p$ and $\hbar \omega$ corresponding
to the first Bogoliubov band. 
In order to detect a sizeable signal at large $p$ and to point out
the corresponding oscillating behaviour of $Z_1(p)$ (see
Figs.\ref{fig_Sp_10}), the intensity $s$ of the optical potential
should be neither too small, nor too large. In fact for $s \to 0$
the strength to the lowest band becomes weaker and weaker if $p$
is outside the first Brillouin zone. For large $s$ the strength
is instead quenched for all values of $p$ because of the presence
of two-body interactions.

In Figs.\ref{fig_Sp_10} we also report the results for the static 
structure factor (full line) corresponding to the sum 
$S(p) = \sum_j Z_j(p)$. One finds that for weak interactions 
(Fig.\ref{fig_Sp_10}(a)) also the static structure factor 
exhibits characteristic oscillations, reflecting the contribution
from the first band. This effect is less pronounced for larger 
values of $gn$ (Fig.\ref{fig_Sp_10}(b)) due to the quenching 
of $Z_1(p)$.
In both cases one observes an important difference with respect 
to the behaviour of $S(p)$ in the uniform gas (\ref{feynman})
(dotted lines in Figs.\ref{fig_Sp_10}).

The behaviour of $S(p)$  at small momenta can be described exactly 
using sum-rule arguments. In fact, phonons exhaust both the non energy
and inverse energy weighted sum rules when $p \to 0$,
high energy bands giving rise to contributions of order $p^2$. 
As a consequence high energy bands contribute to the $f$-sum rule 
(\ref{f-sum-rule}) but can not affect the low $p$ behaviour of the non 
energy weighted moment (\ref{static}) which behaves like $|p|$, nor 
the inverse energy weighted moment (\ref{inv-sum-rule}) which
approaches a constant value when $p \to 0$.
The result is that the low $p$ behaviour of the static structure 
factor is entirely determined by phonon correlations and behaves 
like

\begin{eqnarray}
S(p) 
\raisebox{0.1cm}{$
\;\;\;\; \longrightarrow \;\;\;$}
\hspace*{-1.15cm} 
\raisebox{-0.15cm}{$\; p \to 0 $} \;\;\;
 { |p| \over 2 m^* c}
\label{Sp0}
\end{eqnarray}
consistently with the phononic dispersion law and 
Eqs.(\ref{inv-sum-rule},\ref{inv-sum-rule2})
for the compressibility sum-rule.
It is worth noticing that result (\ref{Sp0}) holds for any value
of $s$ and $gn$.
In the absence of the optical lattice one has $m^*=m$ and $c$
coincides with the Bogoliubov sound velocity $c_B$.
Since one can write $m^*c=\sqrt{m^* \kappa^{-1}}$ and both $m^*$
and $\kappa^{-1}$ increase with $s$, one finds that the presence of 
the lattice results in a suppression of the static structure factor 
at low values of $p$, as clearly shown in Figs.\ref{fig_Sp_10}.

Let us conclude by recalling that in Bragg scattering experiments
one actually measures the imaginary part of the response function
$\chi$ rather than the dynamic structure factor.
The two quantities are related by the equation 
${\rm Im}(\chi) = -\pi (S(p,\omega) - S(-p,-\omega))$.
The subtraction between the two terms can be crucial also at
low temperatures. Actually, due to thermal excitation of phonons,
the dynamic structure factor  exhibits a strong temperature dependence 
when $\hbar \omega < k_BT$, even if $T$ is much smaller than the critical 
temperature for Bose-Einstein condensation. However the difference 
$S(p,\omega) - S(-p,-\omega)$ cancels out most of this temperature
dependence so that the measurement of ${\rm Im}(\chi)$ 
provides reliable information on the zero temperature behaviour
of the dynamic structure factor  \cite{zambelli}.

We would like to thank A.~Smerzi for useful discussions.
This research is supported by the Mi\-ni\-ste\-ro dell'Istru\-zio\-ne, 
dell'Uni\-ver\-si\-t\`a e del\-la Ri\-cer\-ca (MIUR).


\begin{references}
\bibitem{note} In deriving Eq.(\ref{t-b}) we have assumed
$n \partial \delta / \partial n \ll \kappa^{-1}$.
The most general form of the dispersion law, whose derivation
goes beyond the scope of this paper \cite{meret2,smerzi}, does not 
affect the conclusions of the present discussion.
\end{references}
\end{document}